\documentclass[oldversion,psfig]{aa}
\usepackage{graphicx}
\usepackage{txfonts}
\usepackage{natbib}
\usepackage{threeparttable}
\bibpunct{(}{)}{;}{a}{}{,}

\usepackage{times}
\newcommand{\Teff}{\ensuremath{T_{\rm eff}}}                
\newcommand{\logg}{\ensuremath{\log g}}                     
\newcommand{\vsini}{\ensuremath{v\sin i}}                   
\newcommand{\Msun}{\ensuremath{\,{\rm M}_\odot}}            
\newcommand{\kms}{\,km\,s$^{-1}$}                           
\newcommand{\EBV}{\ensuremath{E_{B-V}}}                     
\newcommand{\spd}{{\sc spd}}

\begin{document}

\title{Asiago eclipsing binaries program \\ IV. SZ\,Camelopardalis, a $\beta$\,Cephei pulsator in a quadruple, \\ eclipsing system}

   \author{E.~Tamajo\inst{1,2},
           U.~Munari\inst{2},
           A.~Siviero\inst{3},
           L.~Tomasella\inst{2},
           S.~Dallaporta\inst{4}}

   \offprints{etamajo@phy.hr}

  \institute{Department of Physics, University of Zagreb, Bijeni\v{c}ka cesta 32, 10000 Zagreb, Croatia
      \and INAF Astronomical Observatory of Padova, 36012 Asiago (VI), Italy
      \and Department of Astronomy, University of Padova, 35122 Padova (PD), Italy
      \and ANS Collaboration, c/o Osservatorio Astronomico, via dell'Osservatorio 8, 36012 Asiago (VI), Italy
              }


     \abstract{
              We present a spectroscopic and photometric analysis of the
              multiple system and early-type eclipsing binary SZ\,Cam
              (O9\,IV + B0.5\,V), which consists of an eclipsing SB2 pair of
              orbital period P\,=\,2.7 days in a long orbit ($\sim$55\,yrs)
              around a non-eclipsing SB1 pair of orbital period P\,=\,2.8 days. 
              We have reconstructed the spectra of the individual components
              of SZ\,Cam from the observed composite spectra using the
              technique of spectral disentangling. We used them
              together with extensive and accurate $B$$V$$I_{\rm C}$ CCD
              photometry to obtain an orbital solution.  Our
              photometry revealed the presence of a $\beta$\,Cep variable in
              the SZ\,Cam hierarchical system, probably located within the
              non-eclipsing SB1 pair.  The pulsation period is (0.33265\,$\pm$\,0.00005 )\,days
              and the observed total amplitude in the $B$ band is (0.0105\,$\pm$\,0.0005 )\,mag.
              NLTE analysis of the disentangled spectra provided 
              atmospheric parameters for all three components, 
              consistent with those derived from orbital solution.}
	      {}

\keywords{binary stars -- stars: binaries: eclipsing -- stars: fundamental parameters}

\authorrunning{E.\,Tamajo et al.}
\titlerunning{SZ\,Cam}

\maketitle

\section{Introduction}

In the past several decades, new theoretical models have been established in
order to include a number of physical processes and phenomena, which are
important for understanding the structure and evolution of high-mass stars. 
These processes are convective core-overshooting, semi-convection, rotational
material mixing and mass loss due to stellar winds.  However, empirical
constraints on these features remain of high priority, despite steady
improvement in observational techniques and capabilities
(Hilditch 2004).  Eclipsing binaries of the detached type are a great
resource for obtaining accurate values of stellar masses, radii, and luminosities
(Andersen 1991; Torres et al. 2010).  Given that the component
stars have the same age and initial chemical composition, eclipsing binaries
have been extensively used to test predictions of stellar models (eg. 
Siviero et al. 2004, Tomasella et al.  2008a,b).

SZ\,Cam (=HD\,25638  $V$=6.9 mag) is an early type eclipsing binary
(B0\,II-III + O9.5\,V according to Morgan et al. 1955 and
Budding 1975), which is a member of the very young open cluster
NGC\,1502.  For this cluster, the compilation of literature data by Dias et
al.  (2002) lists an age of 10\,Myr, a distance of 1.0\,kpc, and a radial
velocity of $-$9.7 \kms.  The photometric variability of SZ\,Cam was
discovered by (Guthnick \& Prager 1930), and light curves were presented by
Wesselink (1941), Olsen (1961), Kitamura \& Yamasaki (1971), Polushina
(1977), Chochol (1980), Gorda and Polushina (1987), and Gorda (2000).  Their
observations were carried out by means of photographic or photoelectric
photometry.  The photoelectric observations can attain a high intrinsic
accuracy, but not in the case of open clusters where many other field stars
enter the aperture of the photometer (usually several tens of arcsec wide), and 
even less so in the case of SZ\,Cam that has a similarly bright star (HD
25639), itself member of NGC\,1502 and a variable star, just 18 arcsec away. 
All light curves of SZ\,Cam so far published have been judged by these
authors to be noisier than expected from the high brightness of the target
star.  This was attributed to the perturbation of the photometric measure by
nearby stars, and no investigation of an intrinsic variability has been carried 
out in addition to the eclipse modulation.

\begin{table*} \begin{center}
\caption{Comparison of various orbital solutions for
the eclipsing pair in the quadruple system SZ\,Cam.}
\begin{tabular}{ccccccc}
\hline 
&\\
&Harries et al.\,(1998) & Lorenz et al.\,(1998) & Michalska et al.\,(2007) & Gorda (2008) & Mayer et al.\,(2010) & this paper \\
&\\
$K_1$ (km\,s$^{-1}$) & 225.8\,$\pm$\,3.8 & 180.2\,$\pm$\,2.0 & 181.6 & 192.0\,$\pm$\,2.6 & 189.4\,$\pm$\,1.4 & 185.2\,$\pm$\,2.8 \\

$K_2$ (km\,s$^{-1}$) & 259.1\,$\pm$\,4.0 & 261.2\,$\pm$\,3.8 & 268.2 & 266.4\,$\pm$\,2.5 & 264.1\,$\pm$\,2.5 & 247.0\,$\pm$\,3.6 \\

$M_1$  $(M_{\odot})$ & 18.5\,$\pm$\,0.7  & 15.26\,$\pm$\,0.53 &           & 16.65\,$\pm$\,0.38  &  16.6\,$\pm$\,0.4  & 14.31\,$\pm$\,0.54 \\

$M_2$  $(M_{\odot})$ & 16.1\,$\pm$\,0.6  & 10.75\,$\pm$\,0.44 &           & 12.01\,$\pm$\,0.33  &  11.9\,$\pm$\,0.3  & 10.69\,$\pm$\,0.38 \\

$q={{M_2}/{M_1}}$    & 0.871\,$\pm$\,0.026  &  0.690\,$\pm$\,0.015 & 0.68 & 0.72\,$\pm$\,0.02   &  0.717\,$\pm$\,0.035 & 0.747\,$\pm$\,0.006           
\\

$R_1\,(R_{\odot})$   & 8.5\,$\pm$\,0.3  & 9.01\,$\pm$\,0.91 &           &        &  9.4\,$\pm$\,0.2   & 8.91\,$\pm$\,0.05             \\

$R_2\,(R_{\odot})$   & 7.9\,$\pm$\,0.3  & 6.63\,$\pm$\,0.68 &           &        &  5.4\,$\pm$\,0.2   & 6.70\,$\pm$\,0.12              \\

$T_1\,(K)$           & 29\,725\,$\pm$\,500  &  33\,000 &                &        &  30\,500   & 30\,360\,$^*$                          \\
  
${\Delta}T_2\,(K)$   & 2\,542\,$\pm\,724$   &  4\,950\,$\pm$\,500 &     &        &  5\,200\,$\pm$\,740   & 3\,116\,$\pm$\,255         \\

$\gamma$ (km\,s$^{-1}$)& $-$12.0\,$\pm$\,3.0  &  $-$2.9\,$\pm$\,1.6     &          & $-$10.6\,$\pm$\,2.0  & $-$2.3 &          \\

$l_{3}$$^{**}$ & 27.4$\%$  &  20$\%$ &              &        &  30$\%$   & 25.5\,$\pm$\,0.7$\%$                         \\

  $e$  &  0       & 0          &              &            &  0           & 0                   \\
&\\
\hline
\end{tabular} 

     \begin{tablenotes}
     \item[*] {$*$ assumed from NLTE modeling not constrained by orbital solution (see Table~4).}
     \item[**] {$**$ the value of $l_{3}$ is for the $V$ photometric band. }     

     \end{tablenotes}

\end{center} \end{table*}

SZ\,Cam was first recognized as a spectroscopic binary by Plaskett (1924),
and spectroscopically was found to be a triple stellar system by Mayer et al. 
(1994; 2010). The presence of the third component has been
confirmed by speckle interferometry (Mason et al. 1998). More speckle
measurements were published by Gorda et al. (2007) and
Balega et al. (2007).  These speckle measurements detected an orbital
motion for the tertiary component but one that is extended long enough in time 
to cover only a small arc of the overall orbit. Lorenz et al. (1998) and Michalska
et al.  (2007) have found that the tertiary star is itself a close,
single-lined binary, thus making SZ\,Cam a quadruple system.

The main parameters of the orbital solutions so far published (Harries et
al. 1998, Lorenz et al. 1998, Michalska et al. 2007, Gorda 2008, Mayer et
al. 2010) are based on radial velocities obtained over limited wavelength
ranges (sometimes just a single line, as in the case of H$\alpha$ for
Michalska et al. 2007), sometimes on low S/N spectra or at medium to low resolving
powers. The techniques used to measure the radial velocities varied from
multipeaked cross-correlation functions, to simple multi-Gaussian fitting 
to some individual lines, to spectral disentangling of a single line
(H$\alpha$). The resulting orbital parameters differ from one published
solution to the other, much more than the quoted errors as illustrated in
Table~1.

The motivation for this new effort on SZ\,Cam is multifold: 
($i$) to obtain a complete new set of high accuracy $B$$V$$I_{\rm C}$ photometry with long-
focus CCD imaging, able to accurately split SZ\,Cam from any surrounding
disturbing field star and thus greatly improving on existing photometric
data.  In addition to providing better eclipse light curves, these data are
used to investigate and characterize any intrinsic variability displayed by
any of the four components.  We have indeed discovered that one of them is a
$\beta$\,Cep pulsator. \\
($ii$) to derive the radial velocities of the
components of SZ\,Cam via spectral disentangling of the whole optical
spectrum from REOSC \'{E}chelle high-resolution data.  Without
limiting the RV measurement to just a few lines of assumed Gaussian shape as
happened in previous studies, we are confident the results should gain
in accuracy. \\
($iii$) To derive with NLTE analysis the temperature and gravity
of the stellar components of SZ\,Cam, and compare them with the results of
orbital solution and with expectation from theoretical isochrones.

\section{Observations}

\subsection{Spectroscopy}
 
The spectra of SZ\,Cam were obtained in 2005 and 2011 with the
REOSC \'{E}chelle\,+\,CCD spectrograph on the 1.82\,m telescope operated by
Osservatorio Astronomico di Padova at Mt.\,Ekar (Asiago).  The observing log
is given in Table~2.  A 2-arcsec slit was adopted with fixed E-W
orientation, producing a PSF with an FWHM of 1.75 pixel over the whole
observing campaign, corresponding to a resolving power close to
$R_p${$\sim$}20\,000.  The PSF is measured on the night sky and on comparison
lamp emission lines (and confirmed by analysis of telluric O$_2$ and H$_2$O
absorption lines).  Even if uniformly illuminating the slit, the night sky
and comparison lamp emission lines are considered a fair approximation of
the stellar illumination mode of the slit given the typical seeing (around
2\,arcsec) and the deliberate manual guiding of the star on the slit for
half an hour each exposure.  The manual guiding was preferred over the
equally possible automatic one to avoid introducing spurious, low-amplitude
velocity shifts caused by the star not being kept exactly on the slit center 
by the auto-guiding.

\begin{table}[h!]
    \begin{center}
    \caption{\small 
    Journal of observations. The S/N is computed
    on the continuum at 5870\,{$\AA$}. The timescale is UTC. }
    \begin{tabular}{@{~}c@{~}c@{~}c@{~}r|@{~~}c@{~}c@{~}c@{~}r} \hline 
    \multicolumn{8}{c}{}\\ 
    ID                      & 
    HJD                     & 
    Phase                   & 
    \multicolumn{1}{c}{S/N} & 
    ID                      & 
    HJD                     & 
    Phase                   & 
    \multicolumn{1}{c}{S/N} \\
    &(-2450000)&& \multicolumn{1}{c}{}&&(-2450000)&&\\                              
                 
    \multicolumn{8}{c}{}\\
    40964 & 3314.479  &  0.0511  & 128  &     41729 & 3425.479  &  0.9154  &  55  \\
    40972 & 3314.527  &  0.0333  & 144  &     41730 & 3425.484  &  0.9135  &  53  \\
    41098 & 3373.448  &  0.1977  & 114  &     41732 & 3425.494  &  0.9098  &  62  \\
    41153 & 3392.421  &  0.1664  & 138  &     41733 & 3425.500  &  0.9076  &  56  \\
    41219 & 3393.417  &  0.7973  & 140  &     41734 & 3425.505  &  0.9058  &  69  \\
    41276 & 3394.475  &  0.4052  & 129  &     41736 & 3425.521  &  0.8998  &  58  \\
    41435 & 3411.325  &  0.1608  & 127  &     41737 & 3425.526  &  0.8980  &  55  \\
    41436 & 3411.333  &  0.1578  & 130  &     41738 & 3425.532  &  0.8958  &  62  \\
    41437 & 3411.341  &  0.1548  & 100  &     41744 & 3426.420  &  0.5667  &  35  \\
    41445 & 3411.387  &  0.1378  & 107  &     41745 & 3426.425  &  0.5648  &  33  \\
    41446 & 3411.395  &  0.1348  &  91  &     41746 & 3426.430  &  0.5630  &  32  \\
    41447 & 3411.403  &  0.1318  & 171  &     41748 & 3426.438  &  0.5600  &  43  \\
    41489 & 3412.415  &  0.7568  & 169  &     41749 & 3426.443  &  0.5581  &  34  \\
    41490 & 3412.423  &  0.7538  &  97  &     41750 & 3426.449  &  0.5559  &  43  \\
    41491 & 3412.430  &  0.7513  & 125  &     51551 & 5618.409  &  0.7672  &  80  \\
    41614 & 3415.459  &  0.6287  &  29  &     51552 & 5618.417  &  0.7701  & 100  \\
    41716 & 3425.382  &  0.9513  &  47  &     51643 & 5643.402  &  0.0294  & 107  \\
    41717 & 3425.388  &  0.9491  &  40  &     51693 & 5644.304  &  0.3636  &  91  \\
    41718 & 3425.393  &  0.9473  &  30  &     51745 & 5646.325  &  0.1126  & 171  \\
    41728 & 3425.474  &  0.9172  &  62  &     51787 & 5663.272  &  0.3930  & 169  \\
    \multicolumn{8}{c}{}\\ 
    \hline
    \end{tabular} 
    \end{center}
    \end{table}

    \begin{table}
    \begin{center}
    \caption{\small $B$$V$$I_{\rm C}$ photometric data for SZ\,Cam. HJD =
    heliocentric JD - 2455000 ($^a$). The timescale is UTC. }
    \begin{tabular}{c@{~~}c@{~~}c@{~}c@{~~}c@{~}c@{~~}c@{~}c}
    \hline
    \multicolumn{8}{c}{}\\
    \multicolumn{1}{c}{date}&
    \multicolumn{1}{c}{HJD}&
    \multicolumn{1}{c}{$V$}&
    \multicolumn{1}{c}{err}&
    \multicolumn{1}{c}{$B$$-$$V$}&
    \multicolumn{1}{c}{err}&
    \multicolumn{1}{c}{$V$$-$$I_{\rm C}$}&
    \multicolumn{1}{c}{err}\\
    \multicolumn{8}{c}{}\\
    2011 02 11.120  & 603.620  & 6.938 & 0.008   & 0.421 & 0.007 & 0.527 & 0.004 \\
    2011 02 11.111  & 603.611  & 6.950 & 0.005   & 0.408 & 0.006 & 0.536 & 0.004 \\
    2011 02 11.102  & 603.602  & 6.941 & 0.006   & 0.423 & 0.006 & 0.540 & 0.004 \\
    2011 02 11.090  & 603.590  & 6.939 & 0.005   & 0.431 & 0.005 & 0.523 & 0.006 \\
    2011 02 11.078  & 603.578  & 6.935 & 0.007   & 0.437 & 0.009 & 0.529 & 0.007 \\
    2011 02 11.074  & 603.574  & 6.926 & 0.006   & 0.438 & 0.007 & 0.512 & 0.004 \\
    2011 02 11.070  & 603.570  & 6.943 & 0.006   & 0.419 & 0.006 & 0.542 & 0.004 \\
    2011 02 11.066  & 603.566  & 6.938 & 0.006   & 0.419 & 0.006 & 0.533 & 0.002 \\
    \multicolumn{1}{l}{...}             
                    &          &       &         &       &       &       &       \\
    \multicolumn{8}{c}{}\\
    \hline
    \end{tabular}
     \begin{tablenotes}
     \item [a] \tiny{$a$: Table\,3 is published in its entirety in the electronic 
     edition of A\&A. A portion is shown here for guidance regarding its form and content.}
     \end{tablenotes}
     \vspace{0.1cm}
    \end{center}
    \end{table}

    \begin{figure}
    \begin{center}
    \includegraphics[height=8.8cm,angle=0]{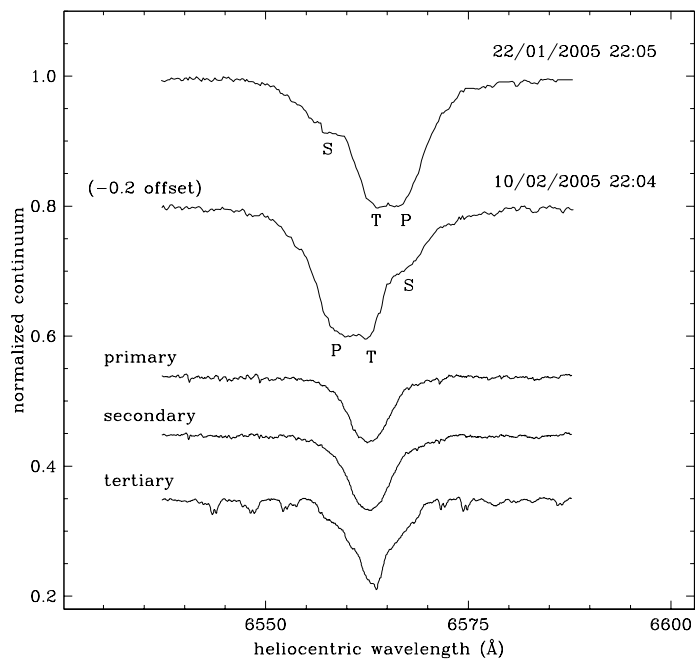}
    \caption{Two examples of H$_\alpha$ profiles for SZ\,Cam from
    Asiago REOSC \'{E}chelle observations.  The features of the
    primary (P), secondary (S) and tertiary (T) stars are labeled.  The
    three lower curves are the spectra of the three individual components,
    normalized to their light contribution, as we disentangled from the
    whole set of 40 REOSC \'{E}chelle spectra covering all orbital
    phases.}
    \label{fig1}
    \end{center}
    \end{figure}

Exposures of a thorium lamp for wavelength calibration were obtained both
immediately before and soon after the exposures on SZ\,Cam, on which the
telescope was still tracking.  These two exposures of the thorium lamp were
combined before extraction, to compensate for spectrograph flexures.  From
the start of the first thorium exposure to the end of the last, the whole
observing cycle including the exposure on SZ\,Cam took about 40\,min. 
According to the detailed investigation and 2D modeling by
Munari \& Lattanzi (1992) of the flexure pattern of the REOSC \'{E}chelle
spectrograph mounted at the Asiago 1.82\,m telescope, and considering that
we imarily observed our target when it was crossing the meridian, the
impact of spectrograph residual flexures on our observations corresponds to
an uncertainty below 0.2\,km/s, thus completely negligible in the
context of our study.  The absence of a systematic velocity off-set and
of random velocity errors lower than 0.3\,km/s are confirmed by ($i$) the
measurement by cross-correlation of the radial velocity of the rich telluric
absorption spectrum in the red portion of all our spectra, and ($ii$) the
measurement of all night sky lines we detected in our spectra, relative to
the compilations of their wavelengths by Meinel et al. (1968),
Osterbrock \& Martel (1992), and Osterbrock et al. (1996).

\subsection{CCD photometry}

CCD photometry on SZ\,Cam was collected in Johnson's $B$ and $V$ bands
and Cousin's $I_{\rm C}$.  A total of 1423 $B$$V$$I_{\rm C}$ observing sequences
were obtained in 43 different observing nights, from August 28, 2010 to
February 11, 2011.  They are reported in Table\,3. Three minima were covered well 
by continuous monitoring throughout the descending, minimum, and ascending branches. 
The time of central eclipse for them occurred at HJD\,462.5310, 535.3861, and 562.3749 (+2455000). 

The telescope was a
0.3\,m f/8 Marcon Richey-Chretien telescope, privately owned by one of us
(S.\,D.) and operated in Cembra (Trento, Italy).  It was equipped with an
SBIG ST-8 CCD camera, 1530\,$\times$\,1020 array, 9~$\mu$m pixels $\equiv$
0.77$^{\prime\prime}$/pix, with a field of view of
19$^\prime$$\times$13$^\prime$.  The $B$$V$$I_{\rm C}$ filters
are from Schuler.

    \begin{figure*}
    \begin{center}
    \includegraphics[height=17.0cm,angle=270]{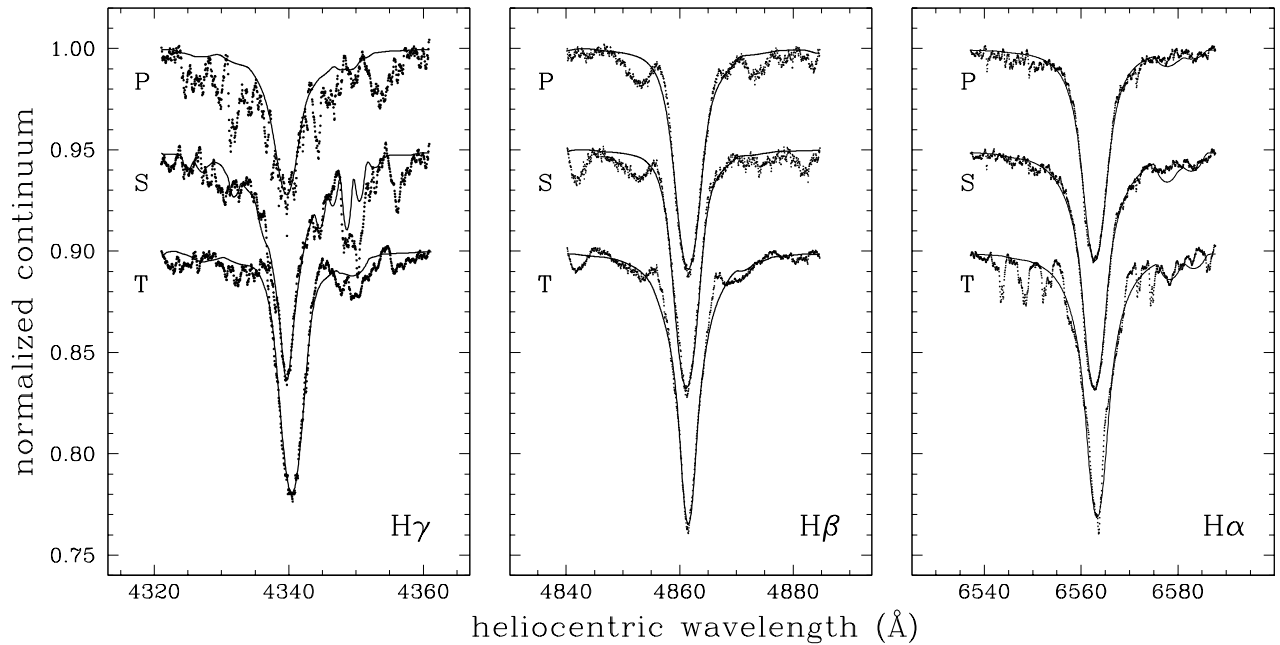}
    \caption{Comparison between the disentangled spectra (points) and
    best-fitting NLTE theoretical spectra (solid lines) for H$_\gamma$, 
    H$_\beta$ and H$_\alpha$ Balmer-line.  Labels P, S, and T mark
    the primary, secondary, and tertiary components, respectively.  }
    \label{fig2}
    \end{center}
    \end{figure*}

The observations were treated in the usual fashion for bias, dark, and
flat-field frames collected in each of the nights when
SZ\,Cam was observed.  The photometric measurements were performed with
aperture photometry, with optimally chosen values for radius of aperture
and internal and external radii of sky annulus.  The optimal values were
automatically and iteratively set by the condition of minimizing
the rms of the many local photometric standard stars on each observation (see next
paragraph) from the linear color equations. In spite of being located within
the well-populated open cluster NGC\,1502, the accuracy of aperture
photometry was not affected by the presence of nearby cluster members, as
extensive tests with PSF fitting photometry proved.  The long focal length,
good seeing, accurate auto-guiding, and the excellent optical quality of the
Richey-Chretien telescope were all instrumental to achieve this result.

The zero point and color equations of the photometry were calibrated
against the $B$,$V$ photometric sequence of Hoag (1961) for NGC\,1502 and
against the $V$$-$$I_{\rm C}$ of the comparison sequence calibrated by
Henden \& Munari (2006) around the nearby CI Cam.  The median slope of the V
color equation was 0.009 with a semi-interquartile range of 0.008.  The
median and the semi-interquartile ranges were 1.008 and 0.017 for the
$B$$-$$V$ color equation, with 0.993 and 0.013 for $V$$-$$I_{\rm C}$.  The
errors reported in Table\,3 are the total budget errors, which are the quadratic
sum of the Poissonian error and the error associated to the transformation to
the standard system as defined by the Hoag (1961) and Henden \& Munari
(2006) standard stars.  The median and semi-interquartile ranges of the total
error budgets for $V$, $B$$-$$V$, and $V$$-$$I_{\rm C}$ are 0.006 and 0.001,
0.007 and 0.002, 0.004 and 0.001 mag, respectively.

\section{Spectral disentangling}

\subsection{The method}

The technique of {\em spectral disentangling} (hereafter {\spd}) allows
isolation of the spectra of the individual component stars of a double-lined
spectroscopic binary system, using as input many observed spectra
distributed well in orbital phases. It was originally formulated in the
wavelength domain by Simon \& Sturm (1994) and in the Fourier domain by
Hadrava (1995).  The technique simultaneously returns the best-fitting
spectra and the orbital velocity amplitudes of the two stars making up the
binary.  A detailed overview of {\spd} can be found in
Pavlovski \& Hensberge (2010).  The disentangled spectrum for each
component of the binary contains the combined signal of all the input
spectra and has a much higher signal-to-noise ratio (S/N) compared to them. 
The total S/N value in a disentangled spectrum of component $n$ is
${S_{dis}\sim {f_n}\,{S_{obs}}\,\sqrt{N_{obs}}}$\,, where $f_n$ is the
fractional light contribution for component $n$ and ${S_{obs}}$ and
${N_{obs}}$ are the S/N and number of input spectra.

The spectra of the individual components can be analyzed in the same way as
those of single stars (Hensberge \& Pavlovski 2007; Pavlovski 2004;
Pavlovski \& Southworth 2009; Pavlovski et al. 2009; Pavlovski \& Hensberge 2005).  The usual
degeneracy between \Teff\ and \logg\, which are  so common in the atmospheric
investigation of isolated single stars, is lifted in the case of eclipsing
binaries by the very precise \logg\ provided directly by the orbital
solution (Hensberge et al. 2000; Simon et al. 1994).

Compared to other methods of radial velocity measurement, {\spd} has some
advantages.  First of all it is independent of template spectra, so it
avoids any systematic errors due to spectral differences between the target
and template stars.  Second, disentangling is not affected by the blending
of spectral lines of the two stars
(see Hensberge et al. 2000; Southworth \& Clausen 2007).  Third, it does not require
the real lines to resemble specific profiles (Gaussians) or to be
symmetric. These two conditions are usually implicitly assumed in many
investigations where blended line profiles are fitted with
combinations of multiple Gaussians.

    \begin{table*} \begin{center}
    \caption{Results of NLTE fitting, with the code
    {\sc GENFITT}, of the Balmer lines on disentangled spectra of SZ\,Cam. 
    The last column gives the reduced $\chi^2$.}
    \begin{tabular}{lrrrrrrrrrrrrc}
    \hline
    &&&&&&&&\\
    line        & \Teff$_1$  & {\logg}$_1$ & {\vsini}$_1$ & ri$_1$ &
                \Teff$_2$  & \logg$_2$ & {\vsini}$_2$  & ri$_2$ & \Teff$_3$  & {\logg}$_3$ & {\vsini}$_3$  & ri$_3$ & $\chi^2_\nu$ \\
                & (K) & [cgs] & (km/s) &   & (K) & [cgs] & (km/s) &  & (K) & [cgs] & (km/s) &  &   \\

    &&&&&&&&\\
	\multicolumn{14}{c}{\it NLTE not constrained by orbital solution}\\
    &&&&&&&&\\

    H$_\alpha$  &  30\,310 & 3.612 & 144.1 & 0.444 & 28\,260 & 3.867 & 117.9 & 0.301 & 26\,635  & 4.223 & 89.9 & 0.255 & 0.745\\  
    & $\pm$190 & $\pm$0.013  & 2.8 & $\pm$0.012 & $\pm$200 & $\pm$0.019 & 3.1 & $\pm$0.014 & $\pm$216 & $\pm$0.031 & 3.7 & $\pm$0.014 &\\
   
    H$_\beta$   &  30\,360 & 3.598 & 143.2 & 0.442 & 28\,420 & 3.793 & 117.1 & 0.304 & 26\,946  & 4.208 & 89.7 & 0.254 & 0.868\\    
    & $\pm$190 & $\pm$0.016  & 3.6 & $\pm$0.017 & $\pm$220 & $\pm$0.028 & 3.3 & $\pm$0.015 & $\pm$229 & $\pm$0.036 & 4.5 & $\pm$0.017 &\\
   
    H$_\gamma$  &  30\,400 & 3.586 & 142.9 & 0.396 & 28\,190 & 3.856 & 116.5 & 0.292 & 26\,854  & 4.212 & 88.7 & 0.312 & 1.204\\    
    & $\pm$250 & $\pm$0.036  & 4.2 & $\pm$0.058 & $\pm$300 & $\pm$0.032 & 4.2 & $\pm$0.053 & $\pm$268 & $\pm$0.054 & 5.4 & $\pm$0.024 &\\

    &&&&&&&&\\
	\multicolumn{14}{c}{\it NLTE constrained with $\logg_1$ and $\logg_2$ from orbital solution}\\
    &&&&&&&&\\

    H$_\alpha$  & 30\,225  & 3.714 & 143.5 & 0.431 & 28\,295 & 3.820 & 117.6 & 0.308 & 26\,589 & 4.239 &  89.8 & 0.261 & 0.814 \\  
        & $\pm$205 & fixed & $\pm$2.9 & $\pm$0.015 & $\pm$210 & fixed & $\pm$3.0 & $\pm$0.016 & $\pm$220 & $\pm$0.035 & $\pm$3.8 & $\pm$0.015 & \\

    H$_\beta$   & 30\,195  & 3.714 & 142.9 & 0.433 & 28\,355 & 3.820 & 116.9 & 0.309 & 26\,895 & 4.224 &  89.4 & 0.258 & 0.984 \\  
        & $\pm$220 & fixed & $\pm$3.4 & $\pm$0.018 & $\pm$225 & fixed & $\pm$3.4 & $\pm$0.017 & $\pm$235 & $\pm$0.041 & $\pm$4.6 & $\pm$0.019 &  \\

    H$_\gamma$  & 30\,209  & 3.714 & 142.1 & 0.422 & 28\,265 & 3.820 & 115.2 & 0.305 & 26\,930 & 4.194 &  88.6 & 0.273 & 1.324 \\  
        & $\pm$270 & fixed & $\pm$3.9 & $\pm$0.042 & $\pm$315 & fixed & $\pm$4.1 & $\pm$0.047 & $\pm$276 & $\pm$0.059 & $\pm$5.2 & $\pm$0.032 &  \\
    &&&&&&&&\\
    \hline
    \end{tabular} 

          \begin{tablenotes}
     \item {ri$_1$, ri$_2$ and ri$_3$ are the relative intensities of the three
                 components to the given line (ri$_1$ + ri$_2$ + ri$_3$=1.0). }
     \end{tablenotes}

\end{center} \end{table*}

There are also a few disadvantages of the {\spd} approach. The first of these
is that the continuum normalization of the input spectra has to be very
precise in order to avoid low-frequency spurious patterns in the resulting
disentangled spectra (Hensberge et al. 2008).  Another disadvantage is that
relative continuum light contributions of the two stars cannot be found
using {\spd}, since this information is itself not contained in the observed
spectra, unless a spectrum has been obtained during an eclipse
(Iliji\'{c} et al. 2004).  

The {\spd} can equally treat eccentric and circular orbits. SZ\,Cam has
been proven by previous studies (and confirmed by our photometric solution
below) to be characterized by zero eccentricity.  To check it further, we
first run the {\spd} by letting unconstrained the eccentricity, and got no
significant deviation from $e$\,=\,0.  Consequently, we fixed a zero
eccentricity and ran a final {\spd} that provided the results used in this
paper.  In setting $e$\,=\,0, the {\spd} converges on the spectra of the
individual components that provide the best fit to a sinusoidal radial
velocity variation.  Therefore, its output amplitude is due to the orbital
velocity variation (and its error).

To characterize the component stars of the binary, we used the method of
fitting synthetic NLTE spectra to the disentangled ones, where the
atmospheric parameters and the projected rotational velocities are
determined simultaneously with the relative light contributions of the
stars.  A genetic algorithm (GENFITT, Tamajo 2009) is used for the
optimization (Tamajo et al. 2011) in order to ensure that the best solution
is found in a parameter space that suffers from strong degeneracies, in
particular between effective temperature (\Teff) and surface gravity
(\logg).

Table~4 compares the results of the
NLTE analysis when it is run un-constrained and when the gravities of the
two stars member of the eclipsing pair are fixed to the values derived
by the orbital solution of Table~5. The rotational velocities of the
tertiary stars are corrected for the smearing introduced by its
orbital motion around an unseen companion.

\subsection{Application to SZ\,Cam}

We analyzed the Asiago REOSC \'{E}chelle spectra of SZ\,Cam
obtained at Mt.\,Ekar about six years apart, using the {\spd} procedure
analysis as implemented in the $FDBINARY$ code (Iliji\'{c} et al. 2004),
including all the 40 observed spectra in Table~2.  We weighted the input
spectra according to their S/N value.

The procedure of {\spd} was performed in several spectral regions.
We converged to accurate results for the regions of the prominent hydrogen 
Balmer lines H$_\alpha$, H$_\beta$, and H$_\gamma$, while other centered on
HeI, HeII, Si\,{II}, and Si\,{III} lines performed less satisfactorily, given the
intrinsic weakness of these lines and the limited S/N of several of our spectra.

The disentangled H$\alpha$ profile for the tertiary component shown in
Figs.~1 and 2 deserves some comments.  The sharp absorptions caused by
telluric O$_2$ and H$_2$O were intentionally {\it not} corrected for in the
input spectra.  The idea was to use them to test the performance of the
{\spd} procedure.  In fact, the wide orbit of the tertiary component around
the eclipsing pair (for which we derive a period of 54.9\,$\pm$\,2.2
\,yr, an eccentricity of the wide orbit of 0.800 \,$\pm$\,0.001 and a time
of periastron passage of 2455343\,$\pm$\,10 \,day, all in good agreement
with previous studies), is so slow that from the point of view of {\spd}
the barycenter of the tertiary component is essentially at rest, when
compared with the large orbital modulation of the component of the eclipsing
pair. The total amplitude of the shift on the telluric lines imposed
by the heliocentric correction is 46\,{\kms}, which is similar to the 50\,{\kms}
amplitude of the orbital motion of the tertiary star around its unseen
companion (Lorenz et al. 1998).  From the point of view of the {\spd}, they
are out of phase with the orbital motion of the main eclipsing pair and
their radial velocity amplitudes are negligible in comparison.  Thus the
only effect impinged on the line profiles of the tertiary star in Fig.~2 is
a moderate broadening of the stellar and telluric lines (the rotational
velocity below derived for the tertiary star is corrected for this
smearing). It is reassuring that the {\spd} procedure attributed the
telluric absorptions to the tertiary component, cleaning them completely
from the disentangled spectra of the primary and secondary components (cf. 
Hadrava 2006).  Particularly interesting in this respect is the presence, on
the H$\alpha$ profile of the tertiary component in Figs.~1 and 2, of the
sharp telluric H$_2$O line at exactly the expected rest wavelength of
6564.21\,${\AA}$, even if this line is well within the H$\alpha$ profile and
just on the red of its photocenter.

     \begin{figure}
     \begin{center}
     \includegraphics[width=8.8cm,angle=0]{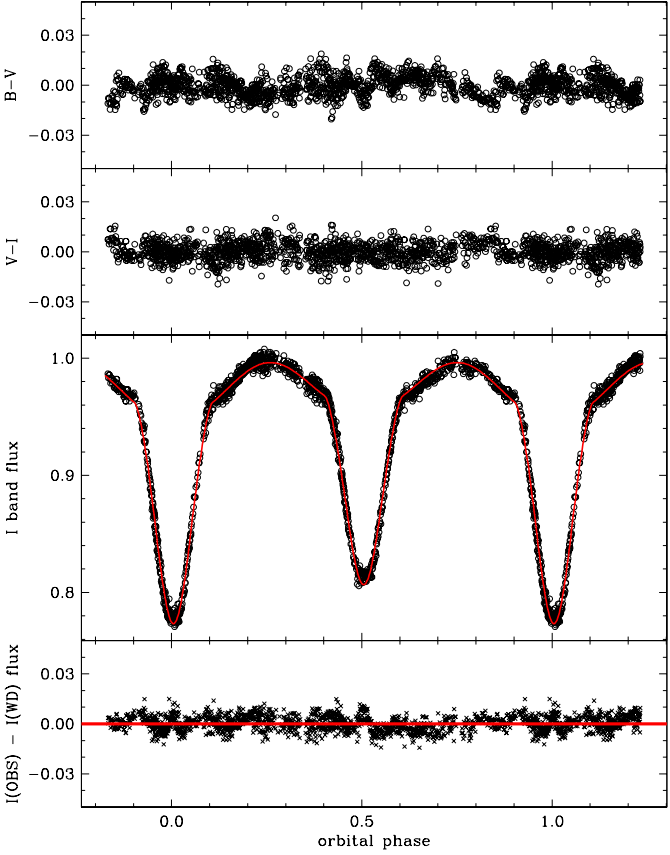}
     \caption{The orbital solution as given in Table~5 is overplotted on
     our photometric data from Table~3. The middle panel shows the $I_{\rm
     C}$ band light curve and the other three panels presents the residuals
     in $(B\,-\,V)$, $(V\,-\,I_{\rm C})$ and $I_{\rm C}$ (top to bottom,
     respectively.}
     \label{fig3}  
     \end{center}  
     \end{figure}

     \begin{figure}
     \begin{center}
     \includegraphics[width=8.8cm,angle=0]{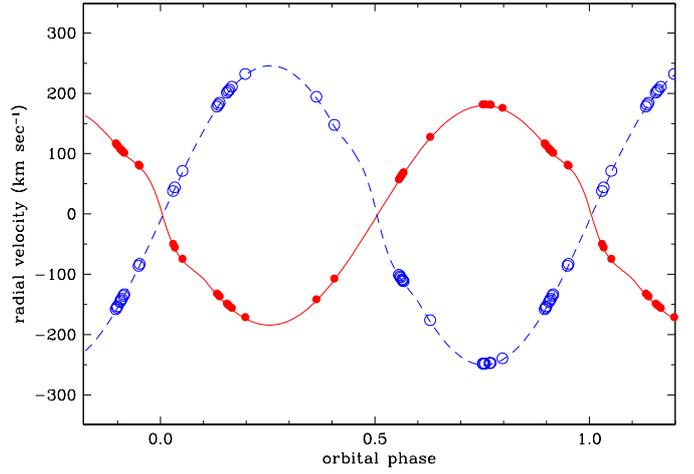}
     \caption{The position in phase of the spectroscopic data used in the
     disentangling is overplotted on the radial velocity curves from the orbital
     solution of Table~5. Filled red circles mark the hotter and more massive 
     (primary) star, and the open blue circles the cooler and less
     massive (secondary) star.}
     \label{fig4}
     \end{center}
     \end{figure}

     \begin{table}
     \begin{center}
     \caption{Light-curve solution, where the given errors are 
      `formal errors' due to the WD solution.} 
     \begin{tabular}{l r c l}
     \hline
     &&&\\
     Parameter&		Value&		&	Error\\
     &&&\\

     {\it period} (days)&		2.6983845&	$\pm$&	0.0000011\\
     $t_\circ$ (primary ecl. HJD)&      2455462.5310&	$\pm$&	0.00023\\ 
     $a$\,(R$_{\odot}$)&		23.97& 		$\pm$&	0.15\\
     $i$\,($^\circ$)&		75.16&		$\pm$&  0.39	\\
     $e$&				0.00&		&	\\
     $T_1$\,(K)&			30\,360&	$\pm$&	(fixed)\\ 
     $T_1 - T_2$\,(K)&		3\,116&		$\pm$&	255\\
     $\Omega_1$&			3.909& 		$\pm$&	0.026\\
     $\Omega_2$&	        	3.592&		$\pm$&	0.025\\
     $R_1$\,(R$_{\odot}$)&		8.91&		$\pm$&	0.05\\
     $R_2$\,(R$_{\odot}$)&		6.70&		$\pm$&	0.12\\
     $M_1$\,(M$_{\odot}$)&	       14.31&		$\pm$&	0.54\\
     $M_2$\,(M$_{\odot}$)&	       10.69&		$\pm$&	0.38\\
     $M_{bol,1}$&			-6.93&		$\pm$&	0.04\\	
     $M_{bol,2}$&			-6.28&		$\pm$&	0.06\\
     log $g_1$\,(cgs)&		         3.70&		$\pm$&	0.01\\
     log $g_2$\,(cgs)&		         3.82&		$\pm$&	0.02\\
     $L_{1}$(Ic band)\,$^a$&         0.444&        	$\pm$&  0.004\\
     $L_{2}$(Ic band)\,$^a$&         0.301&        	$\pm$&  0.005\\
     $l_{3}$~(Ic band)\,$^b$&         0.255&        	$\pm$&  0.010\\
     $R_{1,pole}$\,$^c$  &0.341&      $\pm$&  0.004\\
     $R_{1,point}$\,$^c$ &0.379&      $\pm$&  0.006\\
     $R_{1,side}$\,$^c$  &0.360&      $\pm$&  0.006\\
     $R_{1,back}$\,$^c$  &0.371&      $\pm$&  0.005\\
     $R_{2,pole}$\,$^c$  &0.273&      $\pm$&  0.005\\
     $R_{2,point}$\,$^c$ &0.300&      $\pm$&  0.008\\
     $R_{2,side}$\,$^c$  &0.276&      $\pm$&  0.008\\
     $R_{2,back}$\,$^c$  &0.274&      $\pm$&  0.008\\
     &&&\\
     \hline
     \end{tabular}

     \begin{tablenotes}
     \item[a] \tiny{$a$ relative light contributions ($L_1$\,+\,$L_2$\,+\,$l_3$\,=\,1);}
     \item[b] \tiny{$b$ fractional contribution of the third light at maximum;}
     \item[c] \tiny{$c$ fractional Roche radii in units of separation of mass centers;}                                                                    

     \end{tablenotes}
     \vspace{0.1cm}
     \end{center}
     \end{table}

It is worth noticing that the overall disentangled H$\alpha$ profile for the
tertiary star in Fig.~1 differs significantly from the shape of the same
line for the two other stars. For sake of discussion, we could describe it
as the apparent superposition of a broader and a narrower components. 
Could it be so because our {\spd} procedure has also effectively deconvolved
the two components making up the third star binary?  We plan to
investigate this possibility elsewhere by, in a tailored analysis
combining other spectroscopic and photometric data we are planning to obtain for
this specific purpose.

\section{Orbital solution}

A photometric solution for SZ\,Cam was obtained with the {\sc WD} code
(Wilson \& Devinney 1971; Wilson 1998) in its WD98K93d version as developed by
Milone et al. 1992), by adopting the MODE-2 option, appropriate for detached
binary stars, which implements Roche geometry and a detailed treatment of
reflection and other physical phenomena. It adopted the mass ratio as
provided by the spectral disentangling.  We used the version of the standard
differential corrections procedure ({\sc WDDC}) and iterated until all
parameter corrections were less than their formal errors.  The orbital
solution was then refined by incorporating the information on the radial
velocity semi-amplitudes derived by the spectral disentangling, so that the
radial velocity curve generated by the WD code matched that obtained from
the spectral disentangling.

The full set of fixed and controlled orbital parameters are given in
Table\,5, where the given errors are those formal to the WD solution (the
true errors are probably larger, in particular considering that
the eclipses are only partial and not total). We caution that the formal 
errors can be optimistic when there are strong correlations between parameters,
so should be considered with care (e.g. Southworth, Maxted \& Smalley 2004; Southworth,
Bruntt \& Buzasi 2007). 

The orbital solution is
overplotted to the observed light curve in Fig.\,3 . For our modeling,
we adopted temperature of the primary star as derived by the NLTE analysis,
and the bolometric albedos and gravity brightening exponents appropriate for
radiative atmospheres (Claret 2001, 1998), and limb darkening
coefficients from Van Hamme (1993).  The final solution shows a
negligible dependence on the way limb darkening is accounted for, but it is
very sensitive to the strength of the third light.

The contribution of third light implied by the orbital solution is 25.5\%,
and 26.0\% from NLTE modeling of line-profiles (both methods
to the same wavelength region, the $V$ band, whose effective wavelength is
matched by the weighted average of the NLTE results based on H$_\gamma$,
H$_\beta$ and H$_\alpha$ lines). The two methods therefor converge on the same
contribution of third light within 0.5\%. To test the effect of uncertainty on the amount of 
third light, we reran the WD solution by first increasing the third light for +0.5\% 
and then decreasing it by the same amount (i.e. imposing
$l_{3}$ first to 25.0\% and then to 26.0\%). The effect of such a +0.5\% change in the
contribution of the third light reflects in a change of {\Teff} in the secondary
component by 1.1\%, and a change of 1.2\% in the radii.

Thus, the order of magnitude of the difference in {\Teff} between spectroscopy (NLTE modeling)
and photometry (WD calculation) are a few hundred kelvins. A discussion of the temperature
scale of hot stars is given in Sec. 5.

Our final solution adopted the linear limb darkening law and converged on the
following limb-darkening parameters: x1(bolo)\,=\,0.500, x2(bolo)\,=\,0.500,
and the monochromatic ones as x1(mono)\,=\,0.301, x2(mono)\,=\,0.207 for the
primary and secondary components, respectively.

\subsection{Reddening and distance}

The mean \EBV\ color-excess of NGC\,1502 amounts to about 0.75\,mag according
to Reimann \& Pfau (1987) and \EBV\,=\,0.70 following Dias et al. (2002) with a
possible variation of $\Delta$\EBV\,=\,0.2 over the area of the cluster (Janes
\& Adler 1982, Yadav \& Sagar 2001). The reddening law along the line of
sight to NGC\,1502 deviates from the standard $A_V$\,=\,3.1, because it is shifted 
toward $A_V$\,=\,2.6 (Tapia et al. 1991, Pandey et al. 2003, Weitenbeck et al. 
2008).  Averaging member stars listed by the WEBDA database from the sources of 
$B$,$V$ photometry for NGC\,1502 we built the color-magnitude
diagram of Fig.~6, where the fit is carried out for a Padova isochrone of
solar metallicity and 10\,Myr, scaled in distance and extinction.  The
position of member stars in the diagram is so scattered that the fitting
with an isochrone is not well constrained.  The scatter is intrinsic to the
members of the cluster, as illustrated by the similarly dispersed
color-magnitude diagram published by Michalska et al.  (2009), who
tentatively find \EBV\,=\,0.70 and 1\,kpc distance from their
own photometric data for NGC\,1502.

The equivalent width of interstellar lines correlates with
reddening very well. To check on the reddening to be assumed for SZ\,Cam in computing
its distance, we measured the interstellar lines appearing on our
REOSC \'{E}chelle spectra.  The NaI D1,2 lines are clearly saturated,
indicating a reddening \EBV$\,>\,$0.55, while KI 7699\,${\AA}$ has an
equivalent width of 0.19\,${\AA}$ that corresponds to \EBV\,=\,0.75 following
the calibrations of Munari \& Zwitter (1997).  In the following, we assume
the mean value \EBV\,=\,0.73 as the reddening affecting SZ\,Cam.

To compute a distance to SZ\,Cam from the orbital solution, we adopted
a bolometric correction BC\,=\,$-$3.16 from Bessell et al. (1998) for all
three components, and for the Sun L$_\odot$\,=\,3.826\,x\,10$^{26}$\,W and
M$_{bol,\,\odot}$\,=\,4.74. The classical method of determining the
distance to an eclipsing binary is to calculate the luminosity of each
component star from its radius and effective temperature. The resulting
values of absolute bolometric magnitude, M$_{bol}$, are then converted to
absolute visual magnitudes, M$_{V}$, using bolometric corrections (BCs). 
The combined M$_{V}$ of the components is then compared to the apparent
visual magnitude m$_{V}$ to find the distance modulus (e.g.  Harmanec \&
Pr\v sa 2011; Southworth et al. 2005).

We obtained a distance of 870\,$\pm$\,30 pc for SZ\,Cam.  This lies
comfortably close to the rather uncertain value for the distance of the
parent NGC\,1502 cluster. Unfortunately, this determination cannot be
checked against the Hipparcos parallax for SZ\,Cam, which is useless since 
its error is six times more than the value (Van Leeuwen 2007).

\subsection{$\beta$\,Cep pulsation}

The high accuracy of our photometry allows for searching and characterizing the
low-amplitude intrinsic variability of the individual stars making up SZ\,Cam.
The presence of a stable and periodic variability with a total
amplitude of around 0.01\,mag and a period of about eight hours is already evident
in our data by looking at the individual long photometric observing
runs we collected on SZ\,Cam, which lasted for many consecutive hours.
This microvariability takes the form of a sinusoidal-like disturbance 
around the light curve modulated by eclipses. 

To isolate this variability, we subtracted the flux corresponding to the orbital 
light curve at that phase from each observation, thus working on the
residuals. The residuals were then searched for periodicity using the
Deeming-Fourier algorithm for unequally spaced data (Deeming 1975),
which led again to a very strong peak at 0.33265$\pm 0.00005$ 
days (7$^h$ 59$^m$). The corresponding light curve is presented in Fig.\,5. To plot
the latter, we took the residuals from the orbital light curve for all 1517
observations, divided the 0.33265 day period into 20 bins, and
plotted the mean magnitude and its sigma for each bin. In this way the light curve is
much cleaner than when plotting all 1517 individual points. The
observed total amplitude is $\Delta$B\,=\,0.0105 ($\pm$\,0.0005)\,mag, and
the corresponding ephemeris is \begin{equation} min (B)\,=\,2455311.140(\pm
0.002) + 0.33265 (\pm 0.00005) \times E. \end{equation}

    \begin{figure}
    \begin{center}
    \includegraphics[height=8.8cm,angle=270]{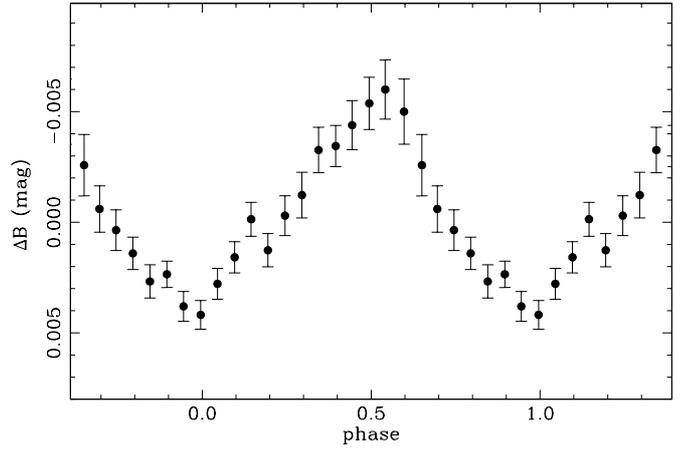}
    \caption{The light curve of the $\beta$\,Cephei star in SZ\,Cam,
    with a period of 0.33265 days and a total amplitude of 0.0105\,mag.}
    \end{center}
    \end{figure}

    \begin{figure}
    \begin{center}
    \includegraphics[height=8.8cm]{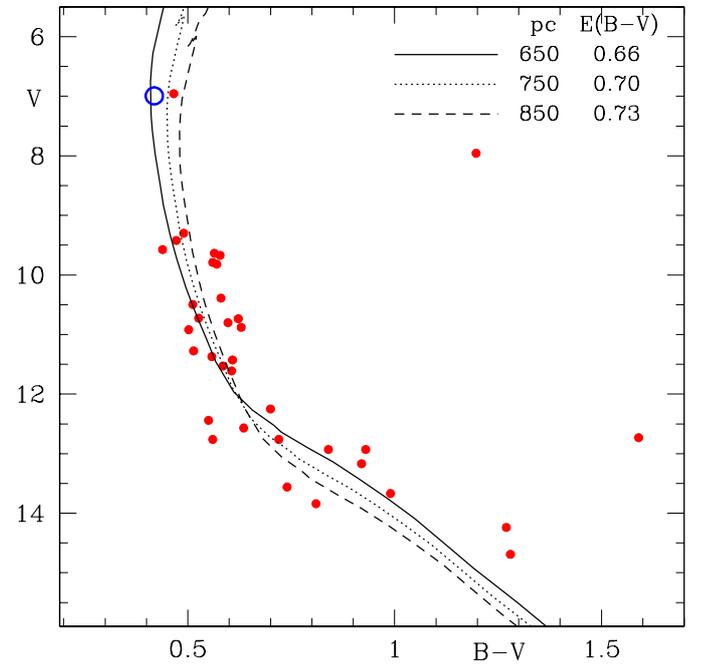}
    \caption{Color magnitude diagram of the open cluster NGC\,1502 built from
    photometric data retrieved from WEBDA
    database (http://www.univie.ac.at/webda/). The blue open
    circle marks SZ\,Cam. The lines are a Padova solar composition isochrone for
    an age of 10\,Myr, shifted according to indicated reddening and distances.}
    \end{center}
    \end{figure}

Such a low-amplitude variability, its sinusoidal light curve, and the length
of the period are typical of $\beta$\,Cep stars. According
to the catalog of $\beta$\,Cep stars of Stankov \& Handler (2005), (hereafter
ST05), their spectral type range from B0 to B3 and the luminosity class from
I to V, with two thirds of the total being composed by B1 and B2 stars of
luminosity classes III, IV, and V.  Their variability is due to pulsations
driven by the $\kappa$-mechanism
(Maeder 2009; Dziembowski \& Pamyatnykh 1993; Sterken \& Jerzykiewicz 1993;
Moskalik \& Dziembowski 1992), and the general aspect of their light curves is
similar to a sinusoid (Hoffmeister 1985; Sterken \& Jaschek 1996).  They are
rare objects, since the ST05 catalog listing only 93 validated $\beta$\,Cep stars. 
The median value of their pulsation amplitude is 0.02\,mag, and the frequency
distribution of the pulsation periods is a broad Gaussian, centered at 0.171
days, with the shortest cataloged period 0.0667 days (V945\,Sco) and
the longest 0.319 days (V595\,Per). According to the ST05 catalog, only
four $\beta$\,Cep stars are eclipsing binaries ($\eta$\,Ori, HD\,92024,
HIP\,84655, and EN\,Lac), and only one of them is also a triple system
($\eta$\,Ori).

In this context, SZ\,Cam stands out as a very interesting member of the
$\beta$\,Cep group of stars, because it has the longest known pulsation period,
it is eclipsing, it is a quadruple system, and it is also member of a young
open cluster. The bottom panel of Figure\,3 suggests that neither at primary
nor at secondary eclipse is the `noise' in the light curve due to the $\beta$\,Cep
pulsation reduced. This seems to indicate that the $\beta$\,Cep variable is 
the primary component of the non-eclipsing pair in the quadruple system
SZ\,Cam. The spectral type of this star (inferred from the other observable
properties) should be B1V, making it a perfect match to the average spectral 
type of known $\beta$\,Cep variables.

\section{Atmospheric analysis and comparison with stellar models}

We performed NLTE analysis of the Balmer lines on the disentangled spectra
of the individual components of SZ\,Cam. These are the strongest lines in
their spectra, and we refrained from using others lines with noisier
profiles because little gain would have been achieved. In OB stars, numerous oxygen, 
nitrogen, and silicon lines, are superimposed on the Balmer-lines are superimposed,
so we carefully selected parts of their profiles that are free of such blendings as
suitable for matching to theoretical line profiles. The code GENFITT has
been developed to fit theoretical line profile to hydrogen lines by
${\chi}^2$ minimization in {\Teff} (Tamajo et al. 2011).

We used a grid of theoretical NLTE spectra that were computed for O-stars
over {\Teff}\,=\,27\,500 - 55\,000\,K in steps of 2\,500\,K and
{\logg}\,=\,3.00 - 4.75\,dex in steps of 0.25\,dex, and for B-stars over
{\Teff}\,=\,15\,000 - 30\,000\,K in steps of 1\,000\,K and {\logg}\,=\,1.75
- 4.75\,dex also in steps of 0.25\,dex.  The results we obtained are
summarized in the upper half of Table~4 (`NLTE not constrained by orbital
- solution') and the theoretical line profile is overplotted to
the observed one in Fig.~2. A warning is necessary here about the reported
results on the third component. The results in Table~4 come from treating
the disentangled line profiles of the third star as if it were a single star. 
However, in discussing Fig.~1 in Sec.~3.2, we consider that our disentangling
could have come close to resolve the tertiary component into the two stars
that compose the SB1 non-eclipsing pair in SZ\,Cam. Therefore, the line we
modeled as coming from a single star could actually be the blend of the
lines of the two components of the SB1 non-eclipsing pair. 

The risk of degeneracy between the atmospheric parameters is always present
when analyzing stellar spectra. To reduce it, we reran the NLTE analysis
by fixing $\logg$ for the primary and secondary star to the value provided
by the orbital solution in Table~5. The results are summarized in the lower
half of Table~4. The differences between the two sets of NLTE results are
minimal, which reinforces confidence in them. The parameter changing the
most is the surface gravity of the primary star, showing a difference of 
mere $\Delta \log g$\,=\,0.1\,dex.

    \begin{figure}
    \begin{center}
    \includegraphics[height=8.8cm,angle=270]{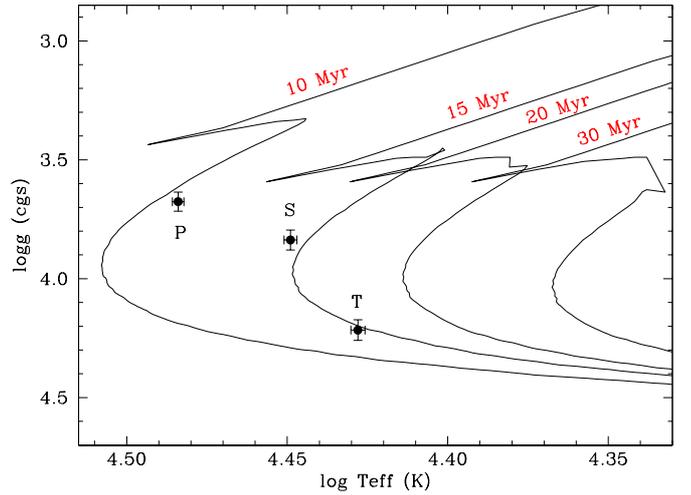}
    \caption{Comparison of the log{\Teff}\,-\,{\logg} plane between the
    observed values for the components of SZ\,Cam and the theoretical
    isochrones of the Geneva group (Meynet et al.  1993), which include the
    effect of stellar rotation and mass loss. Labels P, S, and T mark the
    primary, secondary and tertiary components respectively. The size
    of the error bars corresponds to the formal errors of the orbital
    solution in Table~5.}
    \end{center}
    \end{figure}

    \begin{figure}
    \begin{center}
    \includegraphics[height=8.8cm,angle=270]{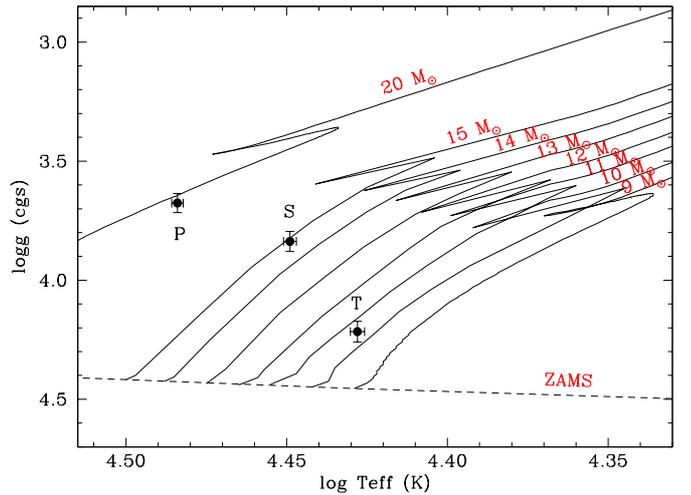}
    \caption{Comparison of the log{\Teff}\,-\,{\logg} plane and the
    observed values for the components of SZ\,Cam and the theoretical
    stellar tracks of the Geneva group (Ekstr\"{o}m et al. 2008), which include the
    effect of stellar rotation and mass loss. Labels P, S, and T mark the
    primary, secondary, and tertiary components respectively. The size
    of the error bars corresponds to the formal errors of the orbital
    solution in Table~5.}
    \end{center}
    \end{figure}

Regarding the effective temperature (\Teff), its determination in OB stars is
a complex task (e.g. B\"{o}hm-Vitense 1981, Crowther 1998). Given the need for a detailed
treatment of non-LTE effects and the presence of
stellar winds (Kudritzki \& Hummer 1990), satisfactory 
modeling of such atmospheres that includes the effects of
numerous metal-lines remains difficult to achieve (cf. Schaerer \& Schmutz 1994, 
Hillier \& Miller 1998, Pauldrach et al. 2001).
Most of published spectral analysis
have so far been based on simple non-LTE models. For
example, the calibration of stellar parameters
of O and early B type stars of Vacca et al. (1996) 
is based only on results from static, plane parallel,
pure hydrogen, and helium (H-He) non-LTE models.
Their derived temperature scale for O stars is found to be
significantly hotter than most earlier calibrations.
Such simplifications non-negligibly affect
the estimated fundamental parameters of O stars (e.g.
luminosities, Lyman continuum fluxes, etc.).
Accurate calibrations of hot stars
are crucial for various astrophysical topics, such as comparisons
with stellar evolution models (and the effects of mass loss and rotation), 
determinations of the initial mass function and ages of clusters, 
ionization balance of HII regions, and supernova progenitors. 

For a given position of OB stars in the HR diagram, the difference in mass between theoretical models
and orbital solutions can be fairly large, for example on the order of 2\,{\Msun} for the binary systems V478\,Cyg
(Popper \& Hill 1991) and  EM\,Car (Andersen \& Clausen 1989) even more than 10\,{\Msun} for the system
HD\,152248 (Penny et al. 1999). Because this involves the transformation from the theoretical HR
diagram (L$_{bol}$ vs {\Teff}) to the observed color-magnitude one (particularly in the $uvby\beta$ system),
some of the discrepancy can be ascribed to uncertainties in involved distance, reddening, 
observed photometry, its transformation to {\Teff} (leading up to $\pm$\,1000-2000\,K uncertainties), 
and the bolometric correction. 

The bolometric correction (BCs) is a very steep function of the
assumed effective temperature ({\Teff}). For O-stars, a 10\%
error in {\Teff} results in an error of 30\% in the derived bolometric
luminosity of a star (see Massey 1998), severely affecting the attempt 
to use stellar evolutionary tracks to determine distances, initial mass functions,
and ages of clusters (see, Massey 1998, 2005; Slesnick et al. 2002). 
In addition, a 10\% uncertainty in {\Teff} results
in a factor of 2 or more uncertainty in the Lyman continuum flux,
affecting our understanding of the ionization balance of HII regions
and the porosity of the interstellar medium in general (see
Oey \& Kennicutt 1997; Oey 2005).

The rotational velocities derived in Table~4 from NLTE analysis of Balmer
lines confirmed the profiles of the SiIII
lines at 4553, 4568, and 4575\,${\AA}$, from which we derived 145\,{\kms},
119\,{\kms}, and 90\,{\kms}, for the primary, secondary, and tertiary
components, respectively. These values are just 1.3\% higher than those
derived from Balmer lines. In view of the relative intensity of SiIII and
Balmer lines, in the rest of this paper we adopt the rotational
velocities from Balmer lines. From the orbital solution in Table~5, the
expected corotation velocities would be 167\,{\kms} for the primary and
126\,{\kms} for the secondary. The secondary essentially match within
errors the condition of corotation, while the primary one, which is already
evolved away from the main sequence, is rotating slower than corotation.

For the third component, assuming it is corotating with the 2.798 day
orbital period of the SB1 non-eclipsing pair, the 89\,{\kms} projected
rotational velocity (cf. Table~4) would correspond to a radius of $R \sim 5
\sin i$ R$_\odot$. This is close to the value of $\sim$5.5~R$_\odot$
expected for a star of the same temperature and luminosity of the tertiary
in SZ\,Cam. This seems to indicate that the orbital inclination for the SB1
non-eclipsing pair is relatively large, too.

Figures~7 and 8 compare the position of the
three components of SZ\,Cam on a temperature-gravity plane with 
Geneva theoretical isochrones from Meynet et al. (1993) 
and stellar tracks from Ekstr\"{o}m et al. (2008), which include mass loss 
and effect of rotation.  These tracks and isochrones,
which are computed for single stars, do not match the properties
of the components of SZ\,Cam as accurately constrained by the analysis
presented in this paper. In Fig.~7 a single isochrone cannot pass through
all three stars at the same time, and the masses inferred by the stellar
tracks in Fig.~8 are much higher than those provided by the orbital solution.
Similar mismatches are common for eclipsing binaries containing young, massive stars,
and are discussed at length in Hilditch (2004).

The evident mismatch probably arises from a combination of stellar models that
would benefit from improvement, and from the effect of SZ\,Cam components
as members of close pairs. Evolution of fast-rotating and heavily mass-
losing stars, trapped in close binary systems, is arguably different from
that of isolated, single counterparts.

    \begin{table}[h!]
    \begin{center}
    \caption{ Summary of the main properties of SZ\,Cam. }
    \begin{tabular}{l@{\hspace*{3pt}}c@{\hspace*{3pt}}c@{\hspace*{3pt}}c@{\hspace*{3pt}}c@{\hspace*{3pt}}}
    \hline
    &&&\\
    Parameter&          system   &        Primary       &        Secondary       &      Tertiary \\
    &&&\\
    E(B-V)         &0.73\,$\pm$\,0.05&&&  \\
    dist.\,(pc)&        870\,$\pm$\,30&&& \\
    $a$ SB2\,(R$_{\odot})$&23.97\,$\pm$\,0.15&&& \\
    mass (M$_{\odot})$&          &   14.31\,$\pm$\,0.54     &  10.69\,$\pm$\,0.38      &                     \\
    radius (R$_{\odot})$&        &    8.91\,$\pm$\,0.05     &   6.70\,$\pm$\,0.12      & (5.0)               \\
    {\Teff} (K)&                 & 30\,320\,$\pm$\,150      &28\,015\,$\pm$\,130       & 26\,760\,$\pm$\,165 \\
    {\logg} (cgs)&               &    3.70\,$\pm$\,0.01     &  3.82\,$\pm$\,0.02       & 4.22\,$\pm$\,0.03 \\
    {\vsini} ({\kms})&           &     145\,$\pm$\,2.3      &    119\,$\pm$\,2.8       &      89\,$\pm$\,3.3     \\
    $M_{bol}$ (mag)&             & $-$6.93\,$\pm$\,0.04     &$-$6.28\,$\pm$\,0.06      &      ($-$5.44)         \\  
    &&&\\
    \hline
    \end{tabular}
    \end{center} 
    \end{table}

The temperatures and gravities of primary and secondary stars are obtained
from a weighted average of values from the orbital solution and the NLTE
analysis. The radius and bolometric magnitude for the third star are estimated under the assumption
that it is corotating in the SB1 non-eclipsing binary (see Sec.\,5). However, see Sec.\,5 for a discussion of   
the difference between method precision and true uncertainties for hot stars.

\section{Summary}

The input data for the analysis presented in this paper were a set of 1517
$B$$V$$I_{\rm C}$ observing sequences obtained in 43 different observing nights,
and 40 high-resolution REOSC \'{E}chelle spectra, well distributed in orbital phase
and varied in S/N.  

In addition to accurately mapping the eclipse light curve, the extreme accuracy of the photometric data, 
for the first time revealed the presence of a
$\beta$\,Cep variable in the SZ\,Cam hierarchical system, probably located
within the non-eclipsing SB1 pair.  The pulsation period is 0.33265 days and
the observed total amplitude in the $B$ band is 0.0105\,mag. The composited REOSC
\'{E}chelle spectra were subjected to spectral disentangling that returned the
individual spectra of the two components of the eclipsing pair and of the
primary of the non-eclipsing SB1 pair.  The disentangled spectra provided
the radial velocity amplitudes of the relative orbital motions, and were
also NLTE-modeled to derive the temperature, surface gravity, and projected
rotational velocities of all components.

This set of data was modeled with the Wilson-Devinney code, and the orbit
of the eclipsing pair was derived. Our orbital solution, while confirming
previous results, significantly improves on the associated formal errors, 
and for the parameters in common it is accurately confirmed by the
results of the NLTE spectral analysis. For the eclipsing pair, the radius
of the primary is 37\% of the orbital separation, 28\% for the secondary.  
Noteworthy is that the relative contribution of the tertiary star to the whole
system brightness turned out to be the same. 

The comparison with theoretical isochrones that incorporates mass-loss and
rotation but which are computed for single, isolated stars, do not match
the properties of the components closely of the SZ\,Cam hierarchical system,
which are evolving in a close binary configuration.  This comparison would
nevertheless suggest that SZ\,Cam is younger than $\sim$20\,Myr, with the
primary and secondary stars well away from the ZAMS, and the tertiary closer
to it.

We hope in the near future to obtain higher resolution and S/N spectra of
this interesting object, with the aim of further constraining the nature and
orbit of the non-eclipsing SB1 pair performing a detailed NLTE chemical
analysis of all its components. The chemical adundances should convincingly
constrain the age and massloss of the components.

\bibliographystyle{aa}

\end{document}